\documentclass[12pt]{iopart}

\usepackage{graphicx}
\usepackage{verbatim}
\usepackage{subfigure}
\usepackage{iopams}
\usepackage[usenames, dvipsnames]{color}
\usepackage{setstack}
\usepackage{amssymb}  
\begin{document}

\title[]{Flux conservation in coherent backscattering and weak localisation of light}

\author{Angelika Knothe and Thomas Wellens}

\address{Physikalisches Institut, Albert-Ludwigs-Universit\"at Freiburg, Hermann-Herder-Str. 3, 79104 Freiburg, D}
\eads{\mailto{Angelika.Knothe@venus.uni-freiburg.de}, \mailto{thomas.wellens@physik.uni-freiburg.de}}

\begin{abstract}
The standard theoretical description of  coherent backscattering, according to which maximally crossed diagrams accounting for interference between counterpropagating path amplitudes are added on top of the  incoherent background, violates the fundamental condition of flux conservation. In contrast to predictions of previous theories, we show that including maximally crossed diagrams with one additional scattering event does not restore flux conservation. Instead, we propose that the latter is recovered when
treating the effects of coherent backscattering and weak localisation in a unified framework. On the basis of this framework, we demonstrate explicitly flux conservation in leading order of the weak disorder parameter $1/(k\ell)$.
\end{abstract}

\pacs{42.25.Dd, 73.20.Fz, 42.25.Fx}


\section{Introduction}

The effect of coherent backscattering leads to an enhancement of intensity in exact backscattering direction when measuring the average wave intensity scattered from a random medium. Known since the years 1984/85 where it has been both described theoretically~\cite{Theo1,Theo2} and observed in experiment~\cite{Ex1,Ex2}, the acknowledged theoretical explanation commonly presented in today's textbooks (see \textit{e.g.}~\cite{MPhEPh}) holds the following mechanism responsible for this phenomenon: The waves propagating within the medium collect random phases due to the random realization of the scattering potential. Thus, upon averaging over different realizations of the disorder, only those constellations of the intensity propagator 
survive  for which the phase difference between the wave and its complex conjugate counterpart vanishes.  Two possible kinds of propagation processes are known to lead to exact phase cancellation: The case where the wave and its complex conjugate visit the same scatterers in the same order (ladder propagation); they propagate the same distance at each step, thereby collecting equal phase. With regard to backscattering, this process leads to a background intensity which is distributed smoothly over all backscattering angles $\vartheta$, since phase cancellation occurs for every backscattering angle (\textit{incoherent background}). 
However, for systems which exhibit reciprocity symmetry \cite{tiggelen97}, exact phase cancellation is also achieved in the case of counter propagating waves (maximally crossed propagation), where again the same scatterers are visited by the wave and the complex conjugate, but in reversed order. Resulting in equal phases in exact backscattering direction,  \textit{i.e.} backscattering angle $\vartheta=0$, but leaving a nonzero random phase shift for every angle other than zero, this gives rise to the sharp backscattering peak centered around $\vartheta=0$, known as the characteristic \textit{coherent backscattering peak}.

This explanation, however, implies a serious deficiency: it is possible to show~\cite{BTh} that for a system where only ladder-like propagation processes are considered, the incoming intensity flux $\varphi_{in}$ of a wave entering a medium is equal to the outgoing intensity flux $\varphi_{out}$, and hence flux conservation is fulfilled. Adding the coherent backscattering peak -- which yields a non-negative
contribution to the outgoing flux for every backscattering angle -- on top of the incoherent background therefore violates the
fundamental law of flux conservation.
Therefore, we may ask which scattering processes counterbalance the maximally crossed contribution and thereby restore flux conservation in the process of coherent backscattering. This question has been addressed before~\cite{Fieb}, where experimental studies are described together with a brief theory to explain the results obtained.

In the work presented here, after briefly outlining the theoretical frame (see chapter~\ref{sec:frame}), we will show that the ansatz proposed in~\cite{Fieb},  \textit{i.e.} including maximally crossed scattering processes which contain one additional scattering event, does not restore flux conservation (see chapter~\ref{sec:scenarios}). Then, we will present the mechanism which we conjecture to meet this goal instead:
  We consider the full set of possible loop propagation processes  by accounting explicitly for the presence of a boundary surface in the case of a finite or semi-infinite medium (see chapter~\ref{sec:wl}). This amounts to a complete and consistent treatment of coherent backscattering together with the effect of weak localisation \cite{chakravarty86} for wave propagation in disordered media. 
  Finally we verify that this consistent treatment indeed restores flux conservation for coherent backscattering in leading order with respect to the disorder strength $1/(k\ell)$ (see chapter~\ref{sec:leading}).

\section{Theoretical frame}
\label{sec:frame}

We treat the propagation of a monochromatic, scalar wave through a disordered medium where it undergoes multiple random scattering. The disorder is described by a potential $V(\mathbf{r})$ denoting a random function of position $\mathbf{r}$. For the disorder average over many realizations of the disorder of any quantity $X$ we write $\overline{X}$. We work in the frame of the Gaussian white noise model being fully characterized by its zero mean value $\overline{V(\mathbf{r})}=0$ and non-zero correlation function $\overline{V(\mathbf{r})V(\mathbf{r}\prime)}= B(\mathbf{r}, \mathbf{r}\prime)$ as well as vanishing cumulants of all orders higher than two. Assuming both, rotational as well as translational invariance of the correlation function, we may write $B(\mathbf{r}, \mathbf{r}\prime)=B(|\mathbf{r}- \mathbf{r}\prime|)$. Furthermore, introducing $\xi_{c}$ as the correlation length of the potential, \textit{i.e.} as the characteristic length of the decay of the correlation function, we restrict ourselves in the following to the regime where the wavelength $\lambda$ of the scattered wave is much larger than the disorder correlation length, \textit{i.e.} to the regime $\lambda \gg \xi_{c}$. In this case, the Dirac delta distribution  is a good approximation for the correlation function for which we hence write 
\begin{equation}
\label{B}
B(|\mathbf{r}- \mathbf{r}\prime|)=u^{2}\delta(\mathbf{r}-\mathbf{r}\prime)\;,
\end{equation}
with $u^{2}$ as a pre-factor taking the role of the scattering strength of the potential. 

To describe the propagation of a wave with wave number $k$ in the presence of the potential $V(\mathbf{r})$ we are in general looking for wave amplitudes $\psi(\mathbf{r})$ satisfying the \textit{scalar Helmholtz equation}
\begin{equation}
\label{Helm}
\left(\Delta+k^{2}\left(1+V(\mathbf{r})\right)\right)\psi(\mathbf{r})=\rho(\mathbf{r})\;,
\end{equation}
for given source distribution $\rho(\mathbf{r})$ where $\Delta=\mathbf{\nabla}^{2}$ denotes the Laplace operator.

For the vacuum case (\textit{i.e.} $V\equiv0$) the Green's function for the Helmholtz operator solving (\ref{Helm}) for a delta-like source $\rho(\mathbf{r})=\delta{(\mathbf{r}-\mathbf{r}\prime)}$ is given by the well-known (retarded) vacuum Green's function~\cite{MPhEPh} 
\begin{equation}
\label{G0}
G^{R}_{0}=G_{0}(\mathbf{r}-\mathbf{r}\prime)=-\frac{e^{ik|\mathbf{r}-\mathbf{r}\prime|}}{4\pi|\mathbf{r}-\mathbf{r}\prime|}
\end{equation}
or by the advanced vacuum Green's function $G_{0}^{A}=(G_{0}^{R})^{*}$ as the complex conjugate of (\ref{G0}). $G_{0}(\mathbf{r}-\mathbf{r}\prime)$ describes the spherical wave at point $\mathbf{r}$ which emanates from a point-like source at $\mathbf{r}\prime$ having propagated the distance $|\mathbf{r}-\mathbf{r}\prime|$ through empty space.

Likewise it is possible to determine the disorder averaged Green's function $\overline{G}(\mathbf{r}-\mathbf{r}\prime)$ to (\ref{Helm}) in the presence of a disorder potential $V(\mathbf{r})\neq0$, where taking the disorder average again implies that instead of specifying the solution for one specific realization of $V(\mathbf{r})$ we describe propagation between the points $\mathbf{r\prime}$ and $\mathbf{r}$ averaged over many different configurations of the disorder. $\overline{G}$ in spatial representation reads~\cite{MPhEPh}
\begin{equation}
\label{Gbar}
\label{Gbarl}
\overline{G}(|\mathbf{r}-\mathbf{r}\prime|) = -\frac{e^{i\tilde{k}|\mathbf{r}-\mathbf{r}\prime|}}{4\pi|\mathbf{r}-\mathbf{r}\prime|},
\end{equation}
and exhibits the same functional form as the vacuum Green's function $G_0$, eq.~(\ref{G0}), except for the wave number $k$ of  (\ref{G0}) which is replaced by $\tilde{k}$ in (\ref{Gbar}). This effective wave number $\tilde{k}$ turns out to be a complex quantity the non-zero imaginary part of which leads to an exponential damping of the Green's function as a consequence of  scattering  within the disordered medium. The corresponding decay constant $l_{s}$ is determined by the imaginary part of $\tilde{k}$ in the following way:
\begin{equation}
\label{ls}
\tilde{k}=k+\frac{i}{2l_s}
\end{equation}
 $l_{s}$ is referred to as the \textit{scattering mean free path} describing the average distance between two successive scattering events.
 For simplicity, we assume that the real part of $\tilde{k}$ is unchanged. (If necessary, this can be achieved by adding a constant term to the potential $V$.)  

 The scattering mean free path certainly depends on the properties of the disordered medium and the scattering processes. In the case of sufficiently dilute media,  \textit{i.e.} if the wavelength of the scattered wave $\lambda=2\pi/k$ is much smaller than the scattering mean free path $l_{s}$, it is justified to assume only the above described ladder propagation processes to survive disorder average. This approximation is called \textit{ladder approximation}.
 On a diagrammatic level, the ladder approximation concerns the irreducible intensity vertex $U$ determining the propagation $\overline{GG^*}$ of the average intensity (see, e.g., \cite{vollhardt80,kuhn07}). However, by means of a Ward identity \cite{vollhardt80} -- which guarantees flux conservation in the case of an infinite medium \cite{kuhn07} -- the intensity vertex is related to the imaginary part of the self energy $\Sigma$ determining the average Green's function $\overline{G}$, see eq.~(\ref{Gbar}), and thereby the scattering mean free path \cite{MPhEPh,Pub}:

\begin{equation}
\label{lsL}
l_{s,Ladder}\equiv\ell=\frac{4\pi}{u^{2}}.
\end{equation}
Hence, within the frame of ladder approximation, the scattering mean free path depends only on the scattering strength $u^{2}$ of the disorder potential as it was introduced in (\ref{B}). 

Corrections to $l_s$ beyond the ladder approximation will be considered in chapter~\ref{sec:leading}.
Furthermore, the expression $k\ell$ turns out to be a system parameter classifying the strength of the disorder by comparing the wavelength of the scattered wave to the scattering mean free path. Hence, a large value of the disorder parameter ($k\ell\gg1\;\iff\;\ell\gg\lambda/2\pi$) signifies weak disorder.

\section{Analysis of different scattering scenarios}
\label{sec:scenarios}

Following the idea of~\cite{Fieb}, according to which the contribution of the maximally crossed scattering sequences ($H_{A}$-contribution in figure \ref{scatt} below) could be cancelled by the contributions of processes which contain one additional scattering event, we analyze in this chapter the contributions of all such scattering sequences one may possibly construct (see figure \ref{scatt}). The choice of the diagrams selected in figure \ref{scatt} can be explained as follows: It is well-known that in the treatment of interference corrections to wave propagation, whenever crossed propagation processes ($H_{A}$) are taken into account, also the corresponding crossed scattering sequences containing one additional scattering event ($H_{B}$ and $(H_{B})^{*}=H_{C}$) play an essential role. This fact manifests itself in the so-called \textit{dressed Hikami-boxes} (see for instance~\cite{MPhEPh}) containing exactly these three scattering processes. It has become clear in earlier works (e. g.~\cite{Kirk1},~\cite{Kirk2}, or~\cite{Pub}), however, that also other scattering processes can play a role of equal importance, i.e. contribute a term with the same asymptotic behaviour for $k\ell\to\infty$ as the $H_{B}$ and $H_{C}$ sequences of the Hikami box. Therefore, figure~\ref{scatt} lists all diagrams which have been identified as the ones giving rise to the leading corrections (scaling like $1/(k\ell)$ or $\ln(k\ell)/(k\ell)^2$  for $k\ell\to\infty$) of the conductivity \cite{Kirk1},~\cite{Kirk2} and the transport mean free path \cite{Pub} (and, related to the latter by means of a Ward identity \cite{vollhardt80}, the scattering mean free path \cite{Pub}) in the case of an infinite medium. Here, we investigate these same processes in the presence of a boundary surface by connecting the respective scattering sequences directly to the incoming and outgoing waves outside the scattering medium, and calculate the corresponding contributions to the backscattered intensity.

\begin{figure}
\centering
\includegraphics[width=100mm]{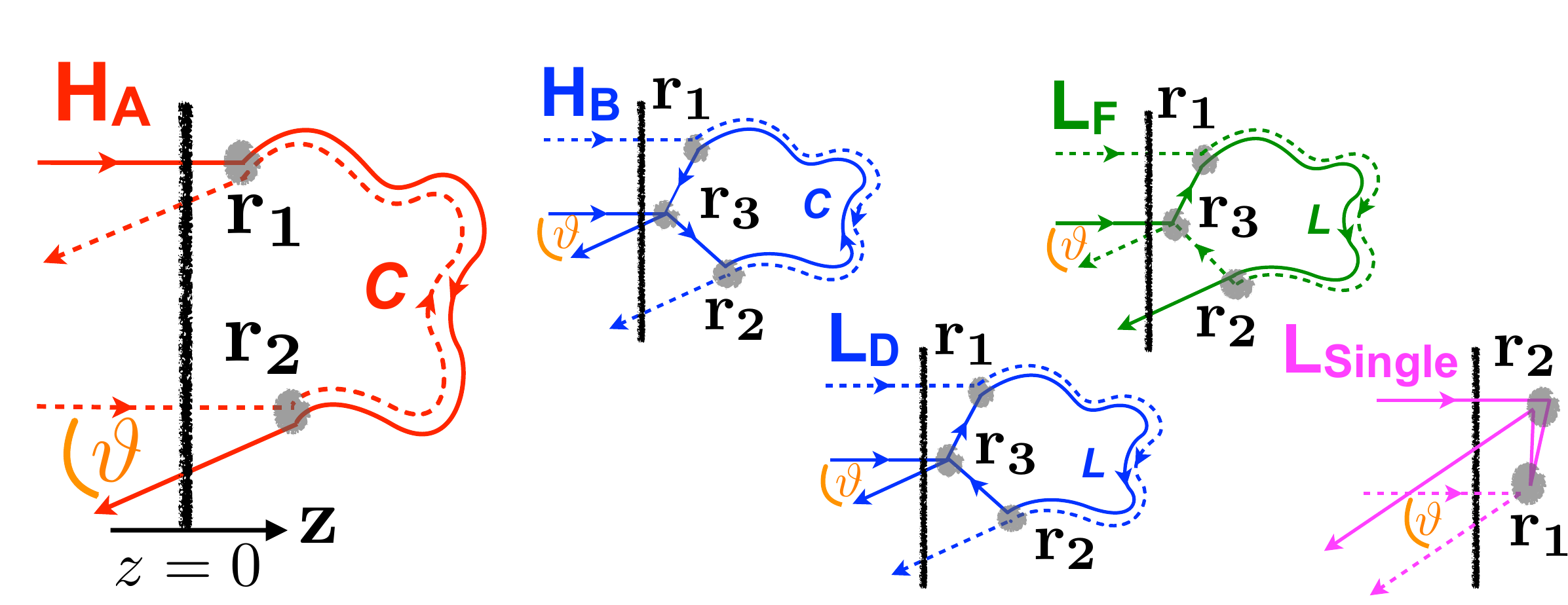}
\caption{Scattering processes investigated beyond ladder approximation in chapter~\ref{sec:scenarios}. 
The choice of these diagrams is motivated by ref.~\cite{Fieb} and previous works on weak localization in infinite media \cite{Pub,Kirk1,Kirk2}.
Solid lines represent Green's functions and dashed lines the respective complex conjugates. The case of maximally crossed propagation ($H_{A}$) is responsible for the famous coherent backscattering cone which, considered separately, violates flux conservation. 
Ref.~\cite{Fieb} claims that $H_{B}$ (together with  $(H_{B})^{*}=H_C$, for complex conjugation exchange every solid line by a dashed line and vice versa) counterbalances the contribution of $H_A$. Here, we additionally take into account $L_{D}$ (or  $(L_{D})^{*}=L_{E}$) -- which, due to the equivalence of ladder and crossed propagation, see (\ref{PC}), yields a contribution identical to $H_B$ -- and the diagrams  $L_{F}$ (or  $(L_{F})^{*}=L_{G}$) and $L_{Single}$ (or  $(L_{Single})^{*}$) -- for which we find a similar asymptotic behaviour for large $k\ell$ as for $H_B$, see eqs.~(\ref{scaling2} -~\ref{scaling4}). As shown in eq.~(\ref{cancel}), however, the total sum of all  these processes does not restore flux conservation. Further diagrams which meet this goal instead are presented in chapter 4.
  }
\label{scatt}
\end{figure}

For the analysis we use the \textit{bistatic coefficient} \cite{ishimaru} $\gamma(\mathbf{\Omega})=\varphi_{out}(\mathbf{\Omega})/\varphi_{in}$ as the ratio of incoming and outgoing intensity flux which is defined as follows:
We assume detection of the scattered intensity at a distance R from the scattering medium sufficiently large to allow far field description and for the emergent wave to be of approximately spherical shape. Therefore, we write
\begin{eqnarray}
\varphi_{out}(\mathbf{\Omega})&=\lim_{R\to\infty}I(R\mathbf{\Omega})4\pi R^{2},\label{flux1}\\
\varphi_{in}&=AI_{0},
\label{flux2}
\end{eqnarray}
where $I(R\mathbf{\Omega})=\overline{|\psi(R\mathbf{\Omega})|^2}$ denotes the average wave intensity at point $\mathbf{R}=R\mathbf{\Omega}$, and $I_0$ and $A$ refer to the intensity of the incoming wave and  the surface of incidence, respectively. Combining eqs.~(\ref{flux1}) and~(\ref{flux2}) we find the bistatic coefficient to read 
\begin{equation}
\label{bist1}
\gamma(\mathbf{\Omega})=\lim_{R\to\infty}\frac{4\pi R^{2}}{AI_{0}}I(R\mathbf{\Omega}),
\end{equation}
Integrating over all angles $\mathbf{\Omega}$, the condition for flux conservation thus reads:
\begin{equation}
\int\frac{d\mathbf{\Omega}}{4\pi}\gamma(\mathbf{\Omega})=1
\label{eq:fluxconsgamma}
\end{equation}
Using this definition, we now calculate the bistatic coefficients for the scattering processes shown in Fig.~\ref{scatt}.
Each diagram contains incoming and outgoing lines associated to the points $\mathbf{r_1}$, $\mathbf{r_2}$ or $\mathbf{r_3}$, described by the functions $\psi_{in}$ and $\overline{G}_{out}$, see eqs.~(\ref{psiin},\ref{GbarlF}) below, and propagators connecting these points with each other, which will be discussed separately below.

\begin{figure}
\centering
\includegraphics[width=25mm]{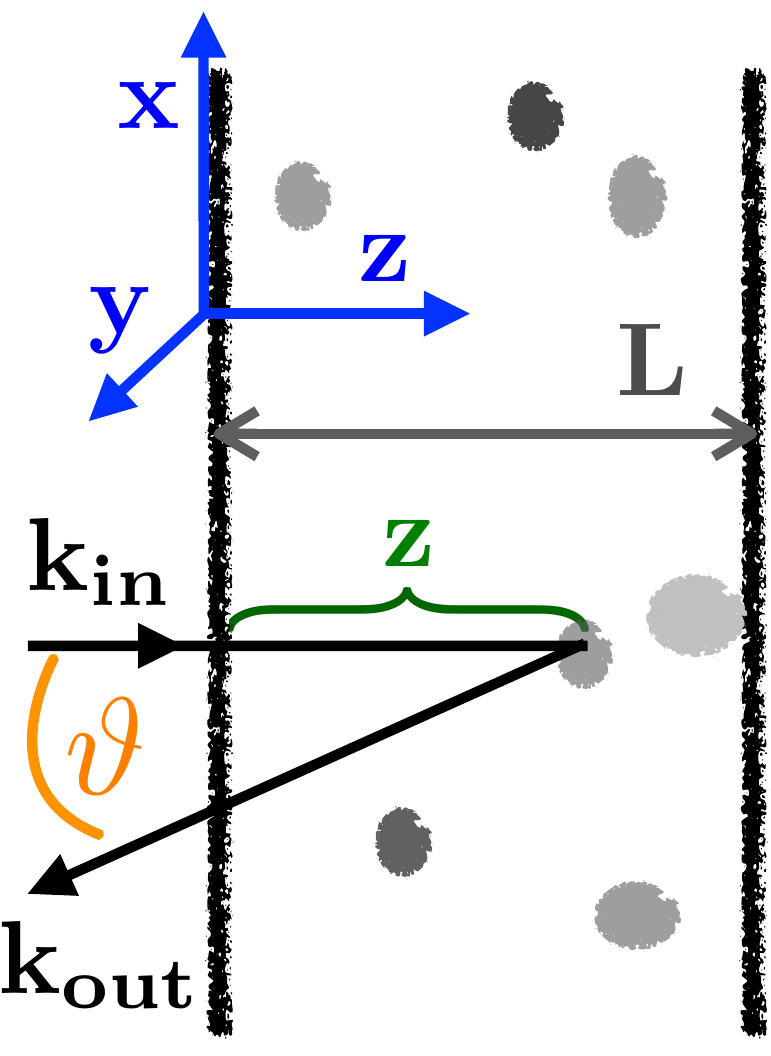}
\caption{Setup and coordinates chosen for the calculations: We consider a slab geometry of finite thickness L in $z$ direction but extended to infinity in the $x$ and $y$ direction. The incoming wave is assumed to be a plane wave propagating in $z$ direction, thus entering the medium perpendicularly to the surface of incidence. The backscattering angle $\vartheta$ is defined as the angle between the incoming and outgoing wave vector: $\vartheta= \sphericalangle (\mathbf{k}_{out},\mathbf{k}_{in})$.}
\label{set}
\end{figure}

For the calculations we consider a slab geometry exhibiting finite thickness $L$ in $z$-direction but extended to infinity in $x$ and $y$-direction (see figure \ref{set}). This implies $A\to\infty$ in (\ref{bist1}), which, however, will be counterbalanced in the following by restricting integrations over the whole
scattering medium to integrations over $z$:
\begin{equation}
\frac{1}{A}\int d\mathbf{r} \rightarrow \int dz.
\end{equation}
Hence, if not explicitly indicated otherwise, the z-integrations are carried out over the half-space of positive z-values within the range of the slab's thickness L, i.e. we assume  $0 < z < L$  in the following. 

The optical thickness of the medium is given by $b=L/\ell$. We assume the incoming wave to be plane with wave vector $\mathbf{k}_{in}=k\mathbf{e}_z$   perpendicular to the $x$-$y$-plane as the surface of incidence at $z=0$:
\begin{equation}
\label{psiin}
\psi_{in}(\mathbf{r})=e^{-\frac{z}{2l_s}}e^{i\mathbf{k}_{in}\cdot\mathbf{r}}=e^{-\frac{z}{2l_s}}e^{ik z},
\end{equation}
The outgoing wave is described by the average Green's function as given in (\ref{Gbarl}) in Fraunhofer approximation~\cite{MPhEPh}: 
\begin{equation}
\label{GbarlF}
\overline{G}_{out}(\mathbf{r}, \mathbf{R})=e^{-\frac{z/\cos\vartheta}{2l_s}}\frac{e^{-ik| \mathbf{R}-\mathbf{r}|}}{4\pi| \mathbf{R-\mathbf{r}}|}\approx e^{-\frac{z/\cos\vartheta}{2l_s}}e^{-i \mathbf{k}_{out}\cdot \mathbf{r}}\frac{e^{ikR}}{4\pi R},
\end{equation}
with $\mathbf{k}_{out}=k\mathbf{R}/R$ and $z/\cos\vartheta$ the distance the wave has to travel from the point $\mathbf {r}$ until leaving the medium. Reflection at the boundary of the scattering medium is neglected for, both, incoming and outgoing waves  (which is appropriate in the case of weak disorder). Due to rotational symmetry around the $z$-axis, the bistatic coefficient for the slab geometry only depends on the backscattering angle $\vartheta$, see figure~\ref{set}, such that the condition of flux conservation, see eq.~(\ref{eq:fluxconsgamma}), turns into $\int_{-1}^{+1}\frac{d\cos\vartheta}{2}\gamma(\vartheta)=1$.

Concerning the remaining propagators in figure~\ref{scatt}, the curly lines connecting the
points $\mathbf{r_1}$ and $\mathbf{r_2}$ represent the crossed (for diagrams $H_A$ and $H_B$) or ladder ($L_F$ and $L_D$) propagator 
$P_C(\mathbf{r_1},\mathbf{r_2})$ or $P_L(\mathbf{r_1},\mathbf{r_2})$ for the average intensity. The ladder propagator fulfills the
following self-consistent integral equation, describing a ladder-like propagation process between the points $\mathbf{r_1}$ and $\mathbf{r_2}$ \cite{Theo2}:
\begin{equation}
\label{PL}
P_{L}(\mathbf{r_1},\mathbf{r_2})=P_{0,L}(\mathbf{r_1},\mathbf{r_2})+\int d\mathbf{r_{i}}P_{L}(\mathbf{r_1},\mathbf{r_{i}})P_{0,L}(\mathbf{r_{i}},\mathbf{r_2}),
\end{equation}
where $\mathbf{r_{i}}$ stands for the positions of all possible intermediate scattering centers we integrate over and $P_{0,L}$ denotes the \textit{single-step ladder propagator} given by
\begin{equation}
\label{PL0}
P_{0,L}(\mathbf{r_1},\mathbf{r_2})=u^{2}\overline{G}(\mathbf{r_1},\mathbf{r_2})\overline{G}^{*}(\mathbf{r_1},\mathbf{r_2})
=\frac{e^{-|\mathbf{r_1}-\mathbf{r_2}|/\ell}}{4\pi\ell |\mathbf{r_1}-\mathbf{r_2}|^2},
\end{equation}
with $l_s=\ell$ in eqs.~(\ref{Gbarl},\ref{ls}). Eq.~(\ref{PL}) can be interpreted as a random walk of a classical particle with average step length $\ell$ and isotropic scattering events between the individual steps. The corresponding  bistatic coefficient results as
\begin{eqnarray}
\gamma_{Ladder}(\vartheta) & = & 4\pi R^2\left (\int dz |\psi_{in}(z)|^2 u^2 |\overline{G}_{out}(\mathbf{r},\mathbf{R})|^2\right.\nonumber\\
& & + \left.\int d z_1d\mathbf{r_2} |\psi_{in}(z_1)|^2 P_L(\mathbf{r_1},\mathbf{r_2}) u^2 |\overline{G}_{out}(\mathbf{r_2},\mathbf{R})|^2\right)\label{eq:ladder}
\end{eqnarray}
(with the first term describing single scattering) and fulfills flux conservation, i.e. $\int_{-1}^{+1}\frac{d\cos\vartheta}{2}\gamma_{Ladder}(\vartheta)=1$ \cite{BTh}. Hence, all additional processes added on top of the incoherent ladder background $\gamma_{Ladder}(\vartheta)$ have to cancel each other mutually in order to maintain flux conservation.

The crossed propagators in figure~\ref{scatt} describes a process where the two conjugate amplitudes propagate in different directions (one from  $\mathbf{r_1}$ to $\mathbf{r_2}$, the other one from $\mathbf{r_2}$ to $\mathbf{r_1}$). However, since the Green's function is symmetric under exchange of its arguments,
the resulting propagator is equivalent to the ladder propagator
\begin{equation}
\label{PC}
P_{L}(\mathbf{r_1},\mathbf{r_2})=P_{C}(\mathbf{r_1},\mathbf{r_2}),
\end{equation}
We now consider separately the different scattering processes shown in figure~\ref{scatt}:
\begin{description}
\item[ $H_{A}$:] 
Crossed contribution process which is responsible for the effect of coherent backscattering. Considering its contribution, only, entails violation of flux conservation. We write for the corresponding bistatic coefficient
\begin{eqnarray}
\label{gHA}
\nonumber\fl\gamma_{H_{A}}(\vartheta)&=4\pi R^{2}\int\int dz_{1}d\mathbf{r}_{2}\psi_{in}(z_{1})\psi^{*}_{in}(z_{2})P_{H_{A}}(\mathbf{r}_{1},\mathbf{r}_{2})u^{2}\overline{G}_{out}(\mathbf{r}_{2},\mathbf{R})\overline{G}^{*}_{out}(\mathbf{r}_{1},\mathbf{R})\\
\fl&=\frac{1}{\ell}\int\int dz_{1}d\mathbf{r}_{2}e^{-(z_{1}+z_{2})(\frac{1}{2\ell}+\frac{1}{2\ell\cos{\vartheta}})}P_{C}(\mathbf{r}_{1},\mathbf{r}_{2})e^{i(\mathbf{k}_{in}+\mathbf{k}_{out})(\mathbf{r}_{1}-\mathbf{r}_{2})},
\end{eqnarray}
where we used the explicit forms of $\overline{G}_{out}$  and $\psi_{in}$ given in eqs.~(\ref{psiin}) and (\ref{GbarlF}), respectively (with $l_s=\ell$), as well as the fact that the propagator for a crossed propagation sequence $P_{H_{A}}$ may be identified as
\begin{equation}
\label{PHA}
P_{H_{A}}(\mathbf{r},\mathbf{r}\prime)=P_C(\mathbf{r},\mathbf{r}\prime).
\end{equation}

\item[ $H_{B}$:] Describing a scattering sequence of crossed propagation of wave and complex conjugate now supplemented by an additional scattering event with incoming and outgoing wave encounter at position $\mathbf{r}_{3}$:
\begin{eqnarray}
\label{gHB}
\nonumber\fl\gamma_{H_{B}}(\vartheta)&=4\pi R^{2}\int\int\int dz_{1}d\mathbf{r}_{2}d\mathbf{r}_{3}\psi_{in}(z_{3})\psi^{*}_{in}(z_{1})P_{H_{B}}(\mathbf{r}_{1},\mathbf{r}_{2},\mathbf{r}_{3})u^{2}\overline{G}_{out}(\mathbf{r}_{3},\mathbf{R})\overline{G}^{*}_{out}(\mathbf{r}_{2},\mathbf{R})\\ \nonumber
\fl&=\frac{1}{\ell}\int\int\int dz_{1}d\mathbf{r}_{2}d\mathbf{r}_{3}e^{-\frac{(z_{1}+z_{3})}{2\ell}}e^{-\frac{(z_{2}+z_{3})}{2\ell\cos{\vartheta}}}
P_{H_{B}}(\mathbf{r}_{1},\mathbf{r}_{2},\mathbf{r}_{3})e^{i\mathbf{k}_{in}\cdot(\mathbf{r}_{3}-\mathbf{r}_{1})}e^{i\mathbf{k}_{out}(\mathbf{r}_{2}-\mathbf{r}_{3})},\\ \fl
\end{eqnarray}
where $P_{H_{B}}$ denotes the propagator corresponding to this particular scattering sequence,
\begin{equation}
\label{PHB}
P_{H_{B}}(\mathbf{r}_{1},\mathbf{r}_{2},\mathbf{r}_{3})=\overline{G}(\mathbf{r}_{3},\mathbf{r}_{2})u^{2}P_{L}(\mathbf{r}_{2},\mathbf{r}_{1})\overline{G}(\mathbf{r}_{1},\mathbf{r}_{3}).
\end{equation}
In \cite{Fieb}, the authors claim that
this process -- together with its complex conjugate $\gamma_{H_{C}}(\vartheta)=\gamma_{H_{B}}^*(\vartheta)$ -- restores flux conservation, i.e. that it counterbalances the coherent backscattering contribution $\gamma_{H_{A}}(\vartheta)$.

\item[ $L_{F}$:] A scattering process containing one single additional scattering event at point $\mathbf{r}_{3}$ as in the $H_{B}$-case, but now visited once by the wave and once by the complex conjugate:
\begin{eqnarray}
\label{gLF}
\nonumber\fl\gamma_{L_{F}}(\vartheta)&=4\pi R^{2}\int\int\int dz_{1}d\mathbf{r}_{2}d\mathbf{r}_{3}~\psi_{in}(z_{3})\psi^{*}_{in}(z_{1})P_{L_{F}}(\mathbf{r}_{1},\mathbf{r}_{2},\mathbf{r}_{3})u^{2}\overline{G}_{out}(\mathbf{r}_{2},\mathbf{R})\overline{G}^{*}_{out}(\mathbf{r}_{3},\mathbf{R})\\ \nonumber
\fl&=\frac{1}{\ell}\int\int\int dz_{1}d\mathbf{r}_{2}d\mathbf{r}_{3}e^{-\frac{(z_{1}+z_{3})}{2\ell}}e^{-\frac{(z_{2}+z_{3})}{2\ell\cos{\vartheta}}}
P_{L_{F}}(\mathbf{r}_{1},\mathbf{r}_{2},\mathbf{r}_{3})e^{i\mathbf{k}_{in}\cdot(\mathbf{r}_{3}-\mathbf{r}_{1})}e^{i\mathbf{k}_{out}(\mathbf{r}_{3}-\mathbf{r}_{2})},\\ \fl
\end{eqnarray}
with $P_{L_{F}}$ the propagator for this scattering sequence,
\begin{equation}
\label{PLF}
P_{L_{F}}(\mathbf{r}_{1},\mathbf{r}_{2},\mathbf{r}_{3})=\overline{G}(\mathbf{r}_{3},\mathbf{r}_{1})u^{2}P_{L}(\mathbf{r}_{1},\mathbf{r}_{2})\overline{G}^{*}(\mathbf{r}_{2},\mathbf{r}_{3}).
\end{equation}

\item[ $L_{Single}$:] Degenerate version of $H_{B}$ where the points $\mathbf{r}_{1}$ and $\mathbf{r}_{2}$ are merged,  \textit{i.e.} no scattering sequence takes place between these points:
\begin{eqnarray}
\label{gLSingle}
\nonumber\fl\gamma_{L_{Single}}(\vartheta)&=4\pi R^{2}\int\int dz_{1}d\mathbf{r}_{2}\overline{\psi}_{in}(z_{2})\overline{\psi}^{*}_{in}(z_{1})u^{2}\overline{G}^{2}(\mathbf{r}_{1},\mathbf{r}_{3})u^{2}\overline{G}_{out}(\mathbf{r}_{2},\mathbf{R})\overline{G}^{*}_{out}(\mathbf{r}_{1},\mathbf{R})\\
\fl&=\frac{4\pi}{\ell^{2}}\int\int dz_{1}d\mathbf{r}_{2}e^{-\frac{(z_{1}+z_{2})}{2\ell}}e^{-\frac{(z_{1}+z_{2})}{2\ell\cos{\vartheta}}}
\overline{G}^{2}(\mathbf{r}_{1},\mathbf{r}_{3})e^{i\mathbf{k}_{in}\cdot(\mathbf{r}_{2}-\mathbf{r}_{1})}e^{i\mathbf{k}_{out}(\mathbf{r}_{1}-\mathbf{r}_{2})},
\end{eqnarray}
\end{description}
To analyze the respective contributions, we first investigate the dependence on the backscattering angle $\vartheta$ and, second, we perform angular integration to obtain information about the total backscattered flux $\gamma_{tot}=\int_{-1}^{+1} \frac{d\cos\vartheta}{2}\gamma(\vartheta)$. 

To evaluate the contributions from the respective processes we use a numerical Monte-Carlo algorithm simulating the propagation of the wave through the medium as a random walk. A comparable method has been used for example in~\cite{MC} to analyze the backscattering cone for different shapes of the scattering medium. In all average Green's functions occurring in the above expressions, we use the ladder approximation $l_s=\ell$ for the scattering mean free path. (Weak localisation corrections to $l_s$ will be considered in chapter~\ref{sec:wl} below.)

\begin{figure}
\centering
\subfigure[]{\includegraphics[width=70mm]{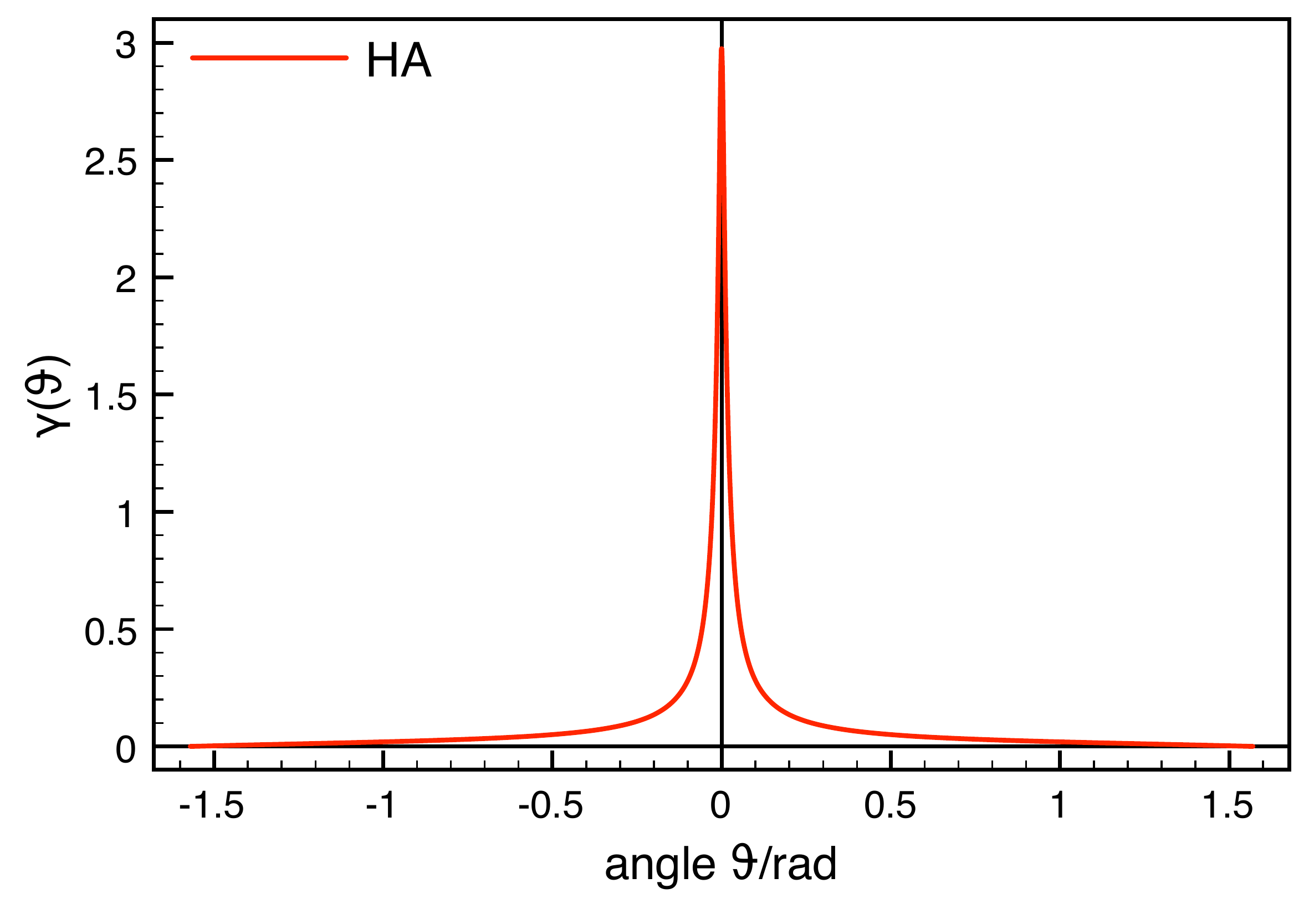}}
\subfigure[]{\includegraphics[width=70mm]{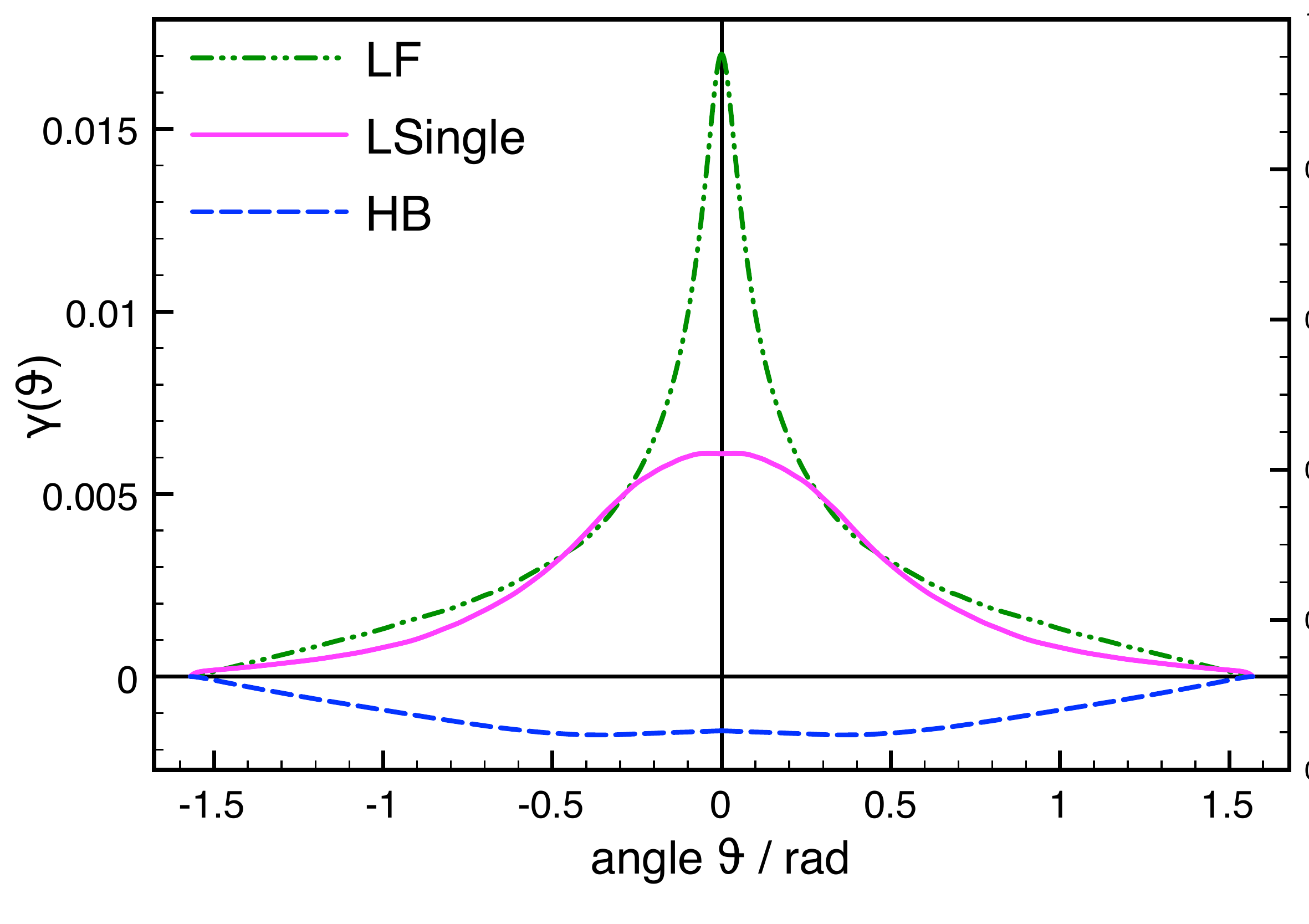}}
\caption{Angle resolved backscattered flux $\gamma(\vartheta)$, see eq.~(\ref{bist1}), for the scattering processes shown in figure \ref{scatt}. System parameters: $k\ell = 30$ and $b=10$.  (a) Characteristic sharp coherent backscattering cone $\gamma_{H_{A}}(\vartheta)$, see eq.~(\ref{gHA}). (b) The contributions of all the processes involving one additional scattering event (real parts of 
$\gamma_{H_{B}}(\vartheta)$, $\gamma_{L_{F}}(\vartheta)$ and $\gamma_{L_{Single}}(\vartheta)$, see eqs.~(\ref{gHB},~\ref{gLF},~\ref{gLSingle})) are found to be several orders of magnitude smaller in height and wider in angular distribution. $H_{B}$ is the only process to contribute negatively and widespread over all angles $\vartheta$ whereas $L_{F}$- and $L_{Single}$-type scatterings yield positive backscattering cones centered around $\vartheta = 0$. }
\label{gamma}
\end{figure}

The results of the angle-resolved analysis are shown in  figure \ref{gamma}.
For the $H_{A}$-contribution, we reproduce the characteristic backscattering cone (figure \ref{gamma}(a)). The contributions of the $L_{F}$- and $L_{Single}$-type scattering processes also yield positive peaks centered around $\vartheta=0$, but they turn out to exhibit a much wider angular distribution than the coherent backscattering cone and to be several orders of magnitude smaller in height. The only scattering processes yielding a negative contribution -- and hence being the only candidates for possible mutual cancellations among the different scattering scenarios -- are $H_{B}$-type scatterings which lead to a small cutback for every backscattering angle $\vartheta$ (figure \ref{gamma}(b)).

In order to find out wether these contributions indeed cancel, we investigate the total backscattered flux $\gamma_{tot}$ of the respective contributions for different values of the disorder parameter $k\ell$. From the results obtained numerically, we derive the following asymptotic scaling laws in the limit of large $k\ell$:
\begin{eqnarray}
\gamma_{tot _{ H_{A}}}&\to & (0.42) \frac{1}{k\ell},\label{scaling1}\\
{\rm Re}(\gamma_{tot _ { H_{B}}}) & \to & -(0.17)\frac{\ln{(k\ell)}}{(k\ell)^{2}},\label{scaling2}\\
{\rm Re}(\gamma_{tot _ { L_{F}}}) & \to & (0.19)\frac{\ln{(k\ell)}}{(k\ell)^{2}},\label{scaling3}\\
{\rm Re}(\gamma_{tot _ { L_{Single}}}) & \to & (0.24)\frac{\ln{(k\ell)}}{(k\ell)^{2}}.\label{scaling4}
\end{eqnarray}
Both in figure~\ref{gamma} and in eqs.~(\ref{scaling1}-\ref{scaling4}), we give only the real parts of the respective contributions since the imaginary parts are cancelled when adding contribution from the corresponding conjugate diagram (e.g. $\gamma_{H_C}=\gamma^*_{H_B}$). We find the leading order contribution to coherent backscattering to originate from the $H_{A}$-type scattering processes, whereas all the scattering processes including one additional scattering event scale in higher order of $1/(k\ell)$. In order to gain further insight as to the origin of the leading order contribution we now additionally distinguish
the $H_{A}$-contribution 
 with respect to the number $N$ of scattering events inside the medium: We first look at scenarios, where the photon undergoes a crossed process including a true multiple scattering sequence ($H_{A,N>2}$), and analyze separately the possibility of double scattering, where the photon leaves the medium after but two scatterings ($H_{A,N=2}$) as shown in the top left sketch of figure \ref{ext}.
Thereby, we obtain the following asymptotic behaviour for large $k\ell$:
\begin{eqnarray}
\label{scaling5}
\gamma_{tot _{ H_{A},N=2}}&\to & (0.42) \frac{1}{k\ell},\\
\gamma_{tot_{H_{A},N>2}} & \to & (0.62)\frac{\ln{(k\ell)}}{(k\ell)^{2}}.
\end{eqnarray}
Hence, the double scattering case of the crossed sequence gives the leading order contribution to the backscattered flux \cite{tiggelen95}.
This leading order term of the order $1/(k\ell)$ can thus not be compensated by any scattering scenario including an additional scattering event, since they were all found to scale in higher order of $1/(k\ell)$. Furthermore, we checked for cancellations among the contributions scaling like $\ln{(k\ell)}/(k\ell)^{2}$:
\begin{eqnarray}
\label{cancel}
& & \gamma_{tot _{ H_{A},N>2}}+ \gamma_{tot _ { H_{B}}} +\gamma_{tot _ { H_{C}}}  \to  0.28 \frac{\ln{(k\ell)}}{(k\ell)^{2}}\neq0,\\
& & \gamma_{tot _{ H_{A},N>2}}+2{\rm Re}\left(2\gamma_{tot _ { H_{B}}} + \gamma_{tot _ { L_{F}}} +\gamma_{tot _ { L_{Single}}}\right)  \to   0.80\frac{\ln{(k\ell)}}{(k\ell)^{2}}\neq0,\label{cancel2}
\end{eqnarray}
where (\ref{cancel2}) amounts to the sum of all processes ($H_B=L_D$, $L_F$ and $L_{Single}$) depicted in figure~\ref{scatt} and their complex conjugates. (Only $H_A$ is identical to its own complex conjugate after exchanging $\mathbf{r_1}$ and $\mathbf{r_2}$.)
In both cases, it is not possible to establish any cancellation of the $H_{A,N>2}$-type contribution. Since (\ref{cancel}) corresponds to the the mechanism of cancellation proposed in~\cite{Fieb}, we state here a clear contradiction to the theory stated beforehand.

A possible reason for this discrepancy might be that, in \cite{Fieb}, a diffusion approximation is employed for the ladder propagator $P_L({\bf r}_1,{\bf r}_2)$. Using this approximation, the authors of \cite{Fieb} arrive at the following analytical expressions:
\begin{eqnarray}
\gamma_{H_{A}}^{({\rm diff})}(\theta) & = & \frac{3/2}{\left(\frac{\mu+1}{2\mu}+q\ell\right)^2}\left(\frac{2\mu}{\mu+1}+\frac{1-\exp(-4q\ell/3)}{q\ell}\right)\\
\gamma_{H_B}^{({\rm diff})}(\theta)+\gamma_{H_C}^{({\rm diff})}(\theta) & \simeq & -\frac{14.45}{(k\ell)^2}\frac{\mu}{\mu+1}
\end{eqnarray}
where $\mu=\cos\theta$ and $q=k\sin\theta$. (Note that these expressions differ by a factor $4\pi$ from those of \cite{Fieb} due to a different definition of the bistatic coefficient.) Calculating the total backscattered flux $\gamma_{tot}=\int_0^1 \frac{d\cos\theta}{2}\gamma(\theta)$ for these expressions yields:
\begin{eqnarray}
\gamma_{tot_{H_A}}^{\rm (diff)} & \to & \frac{3}{4}~\frac{\ln(k\ell)}{(k\ell)^2}\\
\gamma_{tot_{H_B}}^{\rm (diff)}+\gamma_{tot_{H_C}}^{\rm (diff)} & \to & -\frac{2.22}{(k\ell)^2}
\end{eqnarray}
As stated in \cite{Fieb}, both fluxes have different signs and (up to a factor $\ln(k\ell)$) scale like $1/(k\ell)^2$, but, obviously, this does not imply that $\gamma_{tot_{H_A}}^{\rm (diff)} +
\gamma_{tot_{H_B}}^{\rm (diff)}+\gamma_{tot_{H_C}}^{\rm (diff)}=0$.
Therefore, according to, both, our numerical approach and the analytical approach based on the diffusion approximation,
the contributions $H_A$, $H_B$ and $H_C$  originating from the Hikami box do not cancel each other.
Eq.~(\ref{cancel2}) indicates, however, that including all other scattering scenarios with similar scaling behaviour is not the right track towards recovering  flux conservation in coherent backscattering, either.

\section{Coherent backscattering and weak localisation}
\label{sec:wl}

As explained before, propagation processes leading to exact phase cancellation between the wave and its complex conjugate counterpart survive disorder average. In an infinite medium, loop propagation processes (i.e. scattering sequences with equal start and end point) leading to this very scenario are known to be responsible for the effect of \textit{weak localisation} in the case of weak disorder: Just as for the ladder and crossed diagrams yielding the background distribution and the coherent backscattering cone, respectively, for a loop the phase shift between the wave and its complex conjugate vanishes both for the case of equal and reversed pathways. Including these interference paths between the wave and its complex conjugate in the description was found to influence propagation properties since it changes the scattering and the transport mean free path~\cite{Pub}.

\begin{figure}
\centering
\includegraphics[width=60mm]{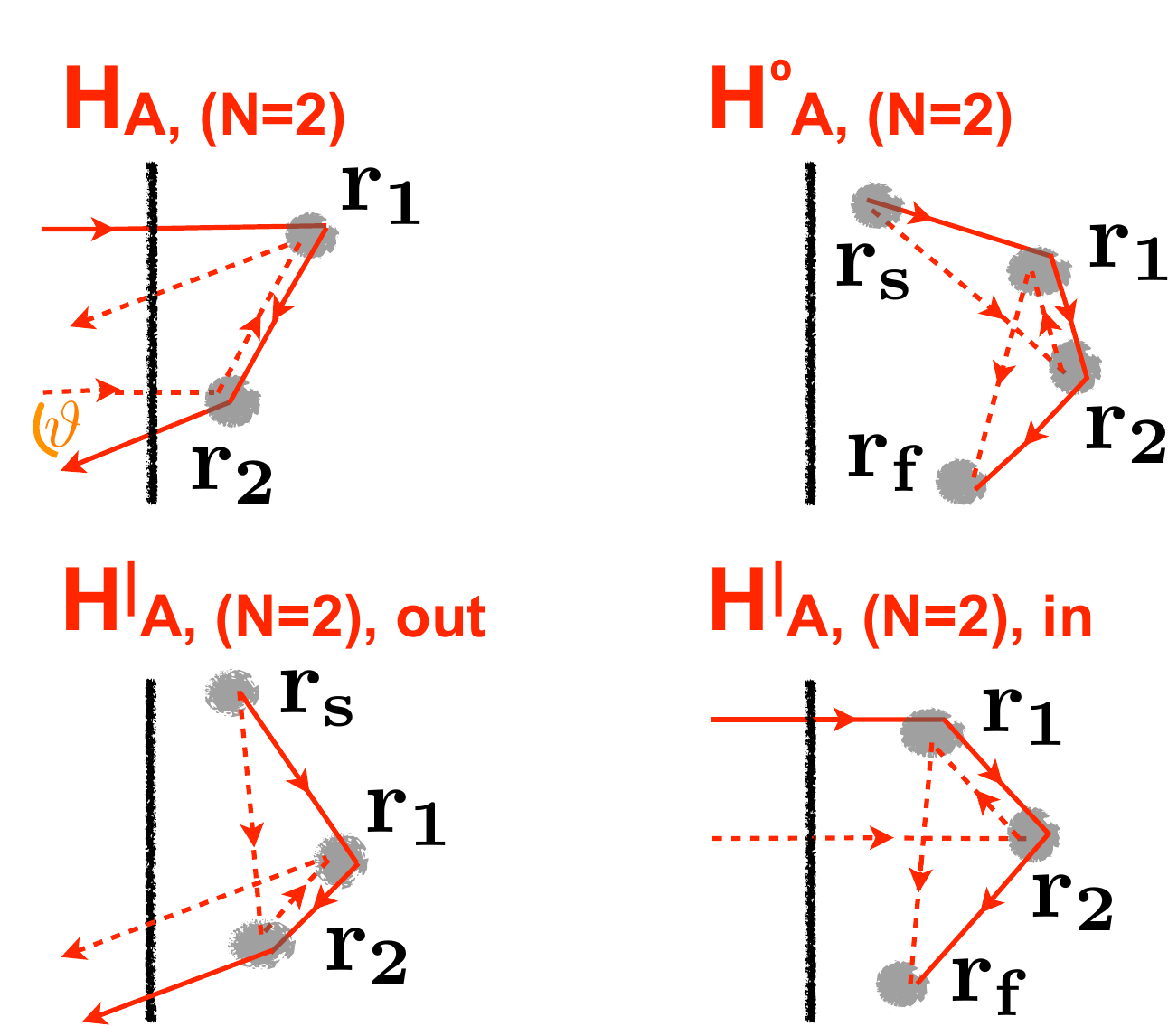}
\caption{Scattering processes taken into account in the frame of a complete description of coherent backscattering (here restricted to double scattering) together with weak localisation. These processes arise naturally when accounting explicitly for the boundary surface of a finite scattering medium:  If the starting point $\mathbf{r}_{s}$ and the endpoint $\mathbf{r}_{f}$ of the propagation are both located inside the medium we deal with a weak localisation loop (top right: $H^{\circ}_{A,N=2}$), whereas the constellation where the end point is located outside the medium while the starting point is still inside (bottom left: $H^{|}_{A, N=2, out}$) contributes to the propagation from $\mathbf{r}_{s}$ to the detector. Similarly, if the endpoint is inside but the starting point outside (bottom right: $H^{|}_{A, N=2, in}$), we obtain a new possibility for the wave to enter and propagate to $\mathbf{r}_{f}$. The coherent backscattering process (top left: $H_{A,N=2}$)  corresponds  to the case were both, the starting and the endpoint, are located outside the scattering medium. The four processes cancel each other mutually, thus ensuring flux conservation.}
\label{ext}
\end{figure}

The top right sketch of figure \ref{ext} ($H^{\circ}_{A,N=2}$) shows the weak localisation  scattering scenario including a crossed propagation process ($H_{A}$) for the case of double scattering ($N=2$). However, if we treat weak localisation in a finite rather than in an infinite medium, now taking into account the presence of the boundary surface, we must allow the possibility for the following processes to occur: Starting and endpoint of the loop (located at points $\mathbf{r_{1}}$ and $\mathbf{r_{2}}$) may also  be located on the other side of the boundary, thus outside the medium (figure \ref{ext} bottom left: $H^{|}_{A,N=2 , out}$ and bottom right $H^{|}_{A, N=2 , in}$, respectively). In the first case, we thus obtain  a further possibility for the wave to leave the medium and propagate to the detector ($H^{|}_{A,N=2 , out}$), and in the second case a new way for the wave to enter the medium ($H^{|}_{A, N=2 , in}$). Of course it is also possible for both, starting and endpoint, to find themselves outside of the scattering medium. By drawing the scattering processes corresponding to this very last scenario, we retrieve the diagram giving rise to the leading coherent backscattering contribution (figure \ref{ext} top left: $H_{A, N=2}).$

Consistency now requires that, as soon as we account for one of the before mentioned propagation processes, we immediately have to consider the other possible constellations as well. This directly leads to a joint treatment of weak localisation and coherent backscattering in the frame of a full treatment of each possible scattering process in the presence of a boundary surface of a finite scattering medium. We will therefore now turn our attention to the analysis of this completed description of coherent backscattering which, as we will see, turns out to be crucial to ensure flux conservation in a finite medium. For this purpose, we will restrict ourselves to $H_{A, N=2}$-type scattering in order to establish a full description of the process yielding the leading contribution to coherent backscattering for weak disorder, as shown in chapter~\ref{sec:scenarios}.

\section{Flux conservation for leading order processes}
\label{sec:leading}

The top right diagram of figure~\ref{ext} gives rise to propagation  from $\bf{r_{s}}$ to  $\bf{r_{f}}$ with intermediate scatterers located at points $\bf{r_{1}}$ and $\bf{r_{2}}$:
\begin{equation}
\label{prop}
P^{\circ}_{H_{A,N=2}}(\mathbf{r_{s}},\mathbf{r_{f}})=u^6\int\int  d\mathbf{r_{1}}d\mathbf{r_{2}} \overline{G}_{s1}\overline{G}^{*}_{s2}|\overline{G}_{12}|^{2}\overline{G}^{*}_{1f}\overline{G}_{2f},
\end{equation}
with $\overline{G}_{12}=\overline{G}(|\mathbf{r_{1}}-\mathbf{r_{2}}|)$,
$\overline{G}_{s1}=\overline{G}(|\mathbf{r_{s}}-\mathbf{r_{1}}|)$, etc.,
 the averaged Green's function as given in (\ref{Gbar}).  
 
For an infinite medium, as shown in \cite{Pub}, this process is associated (by means of a Ward identity) with a change of he scattering mean free path $l_s$ 
such that, in an infinite medium, the sum of $P^{\circ}_{H_{A,N=2}}(\mathbf{r_{s}},\mathbf{r_{f}})$ and the single-step ladder propagator $P_{0,L}(\mathbf{r_{s}},\mathbf{r_{f}})$, see eq.~(\ref{PL0}), is normalized, i.e. $\int d\mathbf{r_f} (P_{0,L}(\mathbf{r_{s}},\mathbf{r_{f}})+P^{\circ}_{H_{A,N=2}}(\mathbf{r_{s}},\mathbf{r_{f}}))=1$ (for an infinite medium).
Equivalently, $l_s$ can also be calculated by means of a certain diagram defining the next-to-leading contributing to the self-energy (see, e.g., \cite{Pub} or exercise 3.8 in \cite{MPhEPh}). In both cases, the result is:
 \begin{equation}
\frac{l_s}{\ell}=1-\frac{\pi}{2k\ell}+\mathcal{O}\left(\frac{\ln(k\ell)}{(k\ell)^2}\right)\label{eq:ls}
\label{cor}
\end{equation}

For our case of a slab geometry, the scattering mean free path may, in principle, differ (at distances of the order of the wavelength from the boundaries) from the scattering mean free path in the infinite medium. However, since $k\ell\gg 1$, this difference can be neglected. Therefore, we will use the scattering mean free path as defined by eq.~(\ref{cor}) also for the semi-infinite medium.

Similarly, the diagrams $H^{|}_{A, in}$ and $H^{|}_{A, out}$ (bottom right and bottom left, respectively), describe the following contributions to the incoming and outgoing wave intensity:
 \begin{eqnarray}
 P^|_{H_{A,N=2,in}}(\mathbf{r_f}) & = & u^4 \int d\mathbf{r_{1}}d\mathbf{r_{2}} 
 \psi_{in}(z_1)\psi_{in}^*(z_2)|\overline{G}_{12}|^{2}\overline{G}^{*}_{1f}\overline{G}_{2f}
 \label{eq:PAin}\\
 P^|_{H_{A,N=2,out}}(\mathbf{r_i}) & = & u^4 \int d\mathbf{r_{1}}d\mathbf{r_{2}} 
\overline{G}_{s1}\overline{G}^{*}_{s2}|\overline{G}_{12}|^{2}\overline{G}_{out}(\mathbf{r_2},\mathbf{R})\overline{G}_{out}^*(\mathbf{r_1},\mathbf{R})\label{eq:PAout}
\end{eqnarray}
with $\psi_{in}$ and $\overline{G}_{out}$ given by eqs.~(\ref{psiin},\ref{GbarlF}). Connecting these building blocks and the single-step ladder propagator with each other,
the total backscattered intensity results as follows: (i) the incoming intensity at point $\mathbf{r}$ is described by the sum of 
the ladder contribution $|\psi_{in}({\bf r})|^2=\exp(-z/l_s)$ (with modified scattering mean free path $l_s$, see eq.~(\ref{eq:ls}) above) and the above contribution
 $P^|_{H_{A,N=2,in}}(\mathbf{r})$. (ii) The intensity then undergoes a random walk consisting of arbitrarily many single steps. Each single step (from  $\bf{r_{s}}$ to  $\bf{r_{f}}$) is described by the sum of the ladder step $P_{0,L}(\mathbf{r_{s}},\mathbf{r_{f}})$ and the weak localization correction $P^{\circ}_{H_{A,N=2}}(\mathbf{r_{s}},\mathbf{r_{f}})$. (iii) From any point $\mathbf{r}_f$, a contribution to the backscattered intensity arises as the sum of the ladder term $\left|\overline{G}_{out}(\mathbf{r},\mathbf{R})\right|^2$ and the new outgoing diagram $P^|_{H_{A,N=2,out}}(\mathbf{r})$. Finally, we add the coherent backscattering cone
 $\gamma_{H_{A,N=2}}(\vartheta)$ on top of the background determined by (i), (ii) and (iii). In total, this gives rise to the following integral equation for the average wave intensity $I(\mathbf{r})=\overline{|\psi(\mathbf{r})|^2}$:
\begin{equation}
I(\mathbf{r})=|\psi_{in}({\mathbf r})|^2+P^|_{H_{A,N=2,in}}(\mathbf{r})+\int d\mathbf{r'}
\left(P_{0,L}(\mathbf{r},\mathbf{r'})+P^{\circ}_{H_{A,N=2}}(\mathbf{r},\mathbf{r'})\right)I(\mathbf{r'})\label{eq:IN2tot}
\end{equation}
from which the outgoing intensity flux results as:
\begin{equation}
\gamma(\vartheta)=\gamma_{H_{A,N=2}}(\vartheta)+\gamma_{Extended}(\vartheta)\label{eq:gammaN2tot}
\end{equation}
where the first term  represents the double scattering contribution to coherent backscattering, whereas the second term amounts to a renormalized background:
\begin{equation}
\gamma_{Extended}(\vartheta)=
4 \pi R^2 u^2 \int dz~ I(\mathbf{r})\left(\left|\overline{G}_{out}(\mathbf{r},\mathbf{R})\right|^2+P^|_{H_{A,N=2,out}}(\mathbf{r})\right)
\label{eq:gammaextended}
\end{equation}
In order to verify that the sum of both terms, eq.~(\ref{eq:gammaN2tot}), fulfills flux conservation, we proceed as follows:
 For $k\ell\gg 1$, it can be shown that 
the main contribution to the double integral over $\mathbf{r_1}$ and $\mathbf{r_2}$ in
eq.~(\ref{prop}) originates from the cases where  $\bf{r_{1}}$ and $\bf{r_{2}}$ are located  very close to each other. We may hence expand the phase-sensitive exponents in the Green's functions occurring  in eq.~(\ref{prop}) in first order of the difference 
$\mathbf{\varepsilon}=\mathbf{r}_{2}-\mathbf{r}_{1}$ and obtain:
\begin{eqnarray}
P^{\circ}_{H_{A,N=2}}(\mathbf{r_{s}},\mathbf{r_{f}}) & \simeq & 
\int d\mathbf{r}d \mathbf{\varepsilon}  
P_{0,L}(\mathbf{r_s},\mathbf{r})P_{0,L}(\mathbf{r},\mathbf{r_f}) u^2 
e^{i\mathbf{q}\cdot\mathbf{\varepsilon}}
\frac{e^{-|\varepsilon|/\ell}}{(4\pi|\varepsilon|)^2}
\nonumber\\
& = & 
\int d\mathbf{r} P_{0,L}(\mathbf{r_s},\mathbf{r})P_{0,L}(\mathbf{r},\mathbf{r_f}) 
\frac{\arctan{q\ell}}{q\ell}\label{Gxp}
\end{eqnarray}
\begin{figure}
 \centering
\subfigure[]{\includegraphics[height=35mm]{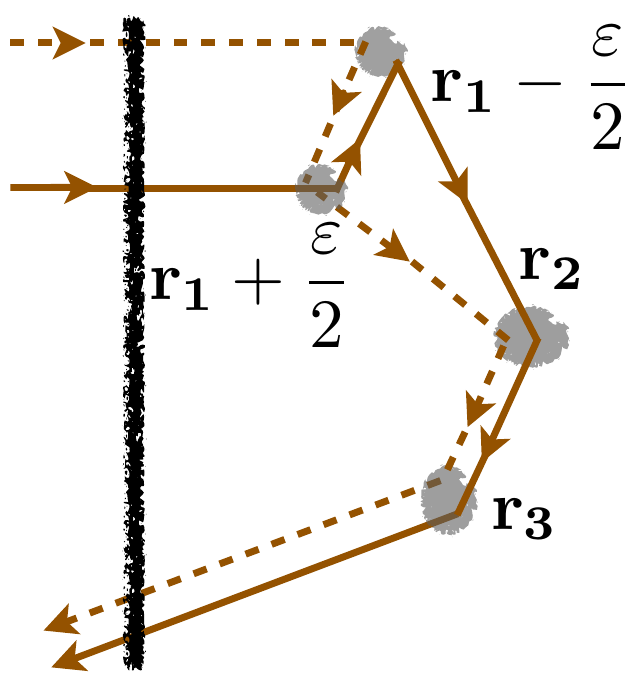}}\hskip10mm
 \subfigure[]{\includegraphics[height=35mm]{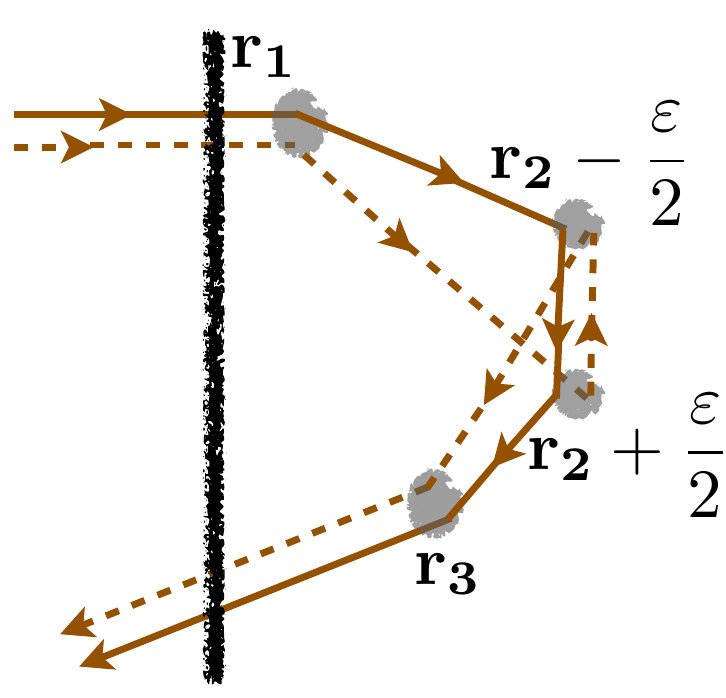}}\hskip10mm
\subfigure[]{\includegraphics[height=35mm]{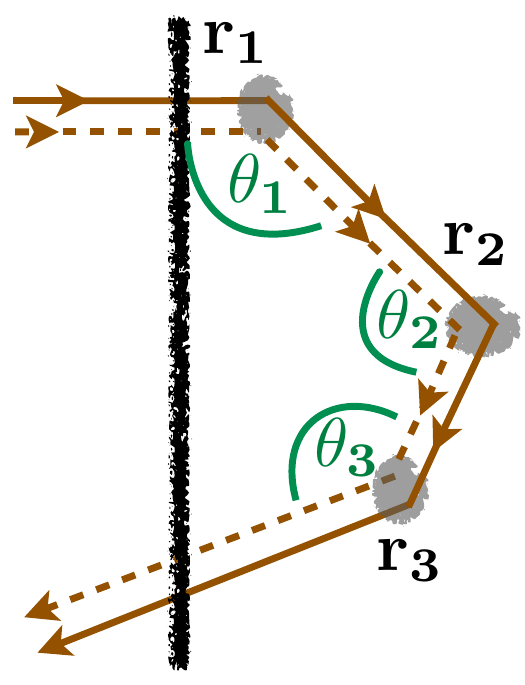}}
\caption{a,b) Two exemplary scattering sequences obtained by combining the building blocks shown in Fig.~\ref{ext} with ladder-like propagation. Sequence a) combines $H^{|}_{A, N=2, in}$, see Fig.~\ref{ext}(bottom right), with an additional ladder step from
$\mathbf{r}_2$ to $\mathbf{r}_3$ and subsequent outgoing ladder propagation. Sequence b) combines $H^{\circ}_{A,N=2}$, see Fig.~\ref{ext}(top right) with incoming and outgoing ladder propagation. c) In the approximate two-step description, see eqs.~(\ref{Gxp}-\ref{eq:gammaHAN2approx}), the sum of all such combinations gives rise to an anisotropic random walk.
According to eq.~(\ref{G}), each scattering event
may be either isotropic (with probability $l_s/\ell$) or anisotropic with distribution $F(\cos\vartheta)$, see eq.~(\ref{eq:Ftheta}). 
For example, sequence a) is reproduced by c) with anisotropic scattering at $\mathbf{r}_1$, and isotropic (i.e. ladder-like) scattering at $\mathbf{r}_2$ and $\mathbf{r}_3$. Likewise, b) corresponds to anisotropic scattering at $\mathbf{r}_2$ in c).
The sum of all these sequences reproduces the leading-order contribution to the coherent backscattering cone $\gamma_{H_{A,N=2}}(\vartheta)$ together with the correspondingly renormalized background, see figure~\ref{ALL}.
}
\label{appr}
\end{figure}
where $\mathbf{r}=(\mathbf{r_1}+\mathbf{r_2})/2$ and $\mathbf{q}=k\left(\frac{\mathbf{r}_{f}-\mathbf{r}}{|\mathbf{r_{f}}-\mathbf{r}|}-\frac{\mathbf{r_s}-\mathbf{r}}{|\mathbf{r_s}-\mathbf{r}|}\right)$. Its absolute value $q=|\mathbf{q}|=k\sqrt{2-2\cos{\vartheta}}$ depends on the angle $\vartheta= \sphericalangle (\mathbf{r}-\mathbf{r_s},\mathbf{r}-\mathbf{r_f})$. Eq.~(\ref{Gxp}) can be interpreted as a two-step ladder propagation
(first from $\mathbf{r_s}$ to $\mathbf{r}$, then from $\mathbf{r}$ to $\mathbf{r_f}$) with non-isotropic intermediate scattering event defined by the angular distribution:
\begin{equation}
\label{dist}
F(\cos\vartheta)=\frac{\arctan{(k\ell\sqrt{2-2\cos{\vartheta}})}}{k\ell\sqrt{2-2\cos{\vartheta}}}.\label{eq:Ftheta}
\end{equation}
Similarly, for $k\ell\gg 1$, the contributions to the incoming and outgoing intensity, eqs.~(\ref{eq:PAin},\ref{eq:PAout}), can be written as  combination of two ladder steps with angle-dependent scattering at $\mathbf{r}$:
\begin{eqnarray}
P^|_{H_{A,N=2,in}}(\mathbf{r_f}) & \simeq & \int d\mathbf{r}
|\psi_{in}(z)|^2 P_{0,L}(\mathbf{r},\mathbf{r_f})F\left(\frac{z-z_f}{|\mathbf{r}-\mathbf{r_f}|}\right)
 \label{eq:PAin2}\\
 P^|_{H_{A,N=2,out}}(\mathbf{r_i}) & \simeq & \int d\mathbf{r} P_{0,L}(\mathbf{r_i},\mathbf{r}) 
 \left|\overline{G}_{out}(\mathbf{r},\mathbf{R})\right|^2 F\left(\frac{(\mathbf{R}-\mathbf{r})\cdot(\mathbf{r_i}-\mathbf{r})}{|\mathbf{R}-\mathbf{r}|~|\mathbf{r_i}-\mathbf{r}|}\right)\label{eq:PAout2}
 \end{eqnarray}
 The same approximation applied to the maximally crossed double scattering contribution yields:
 \begin{eqnarray}
 \gamma_{H_{A,N=2}}(\vartheta) & 
 \simeq &  4\pi R^2 \int_0^L dz~|\psi_{in}(z)|^2 u^2 \left|\overline{G}_{out}(\mathbf{r},\mathbf{R})\right|^2  F(\cos\vartheta)\nonumber\\
 & \simeq & \frac{\cos\vartheta}{1+\cos\vartheta}F(\cos\vartheta)\label{eq:gammaHAN2approx}
 \end{eqnarray}
for $L\gg \ell$.
Integrating eq.~(\ref{eq:gammaHAN2approx}) over all backscattering angles, we recover our previous numerical result, eq.~(\ref{scaling5}), for the asymptotic behaviour $\lim_{k\ell\to\infty} k\ell \int_{0}^{1}\frac{d\cos\vartheta}{2}\gamma_{H_{A,N=2}}(\vartheta)=\pi (\sqrt{2}-\sinh^{-1} 1)/4\simeq 0.42$.

Using eqs.~(\ref{Gxp}-\ref{eq:gammaHAN2approx}), the transport process described by eqs.~(\ref{eq:IN2tot},\ref{eq:gammaN2tot}) reduces to an anisotropic
random walk of a classical particle, see figure~\ref{appr}. For such anisotropic scattering processes, flux conservation is naturally ensured. At each scattering event, the particle may either be scattered isotropically (as it is the case if only ladder diagrams are considered) or anisotropically according to the angular distribution, eq.~(\ref{eq:Ftheta}). (More precisely, eqs.~(\ref{eq:IN2tot},\ref{eq:gammaN2tot}) do not account for sequences with two subsequent anisotropic scattering events which, however, are negligible for $k\ell\gg 1$.) Adding both cases, the total angular distribution results as:
\begin{equation}
\label{G}
G(\cos{\vartheta})=\frac{l_{s}}{\ell}+F(\cos{\vartheta})
\end{equation}
Using eq.~(\ref{eq:ls}), it turns out that  the total distribution is indeed normalized, i.e. $\int_{-1}^{+1}\frac{d\cos\vartheta}{2}G(\cos\vartheta)=1$. (As mentioned above, this is not a coincidence, but can be traced back to a Ward identity.)

\begin{figure}
\centering
\subfigure[]{\includegraphics[height=63mm]{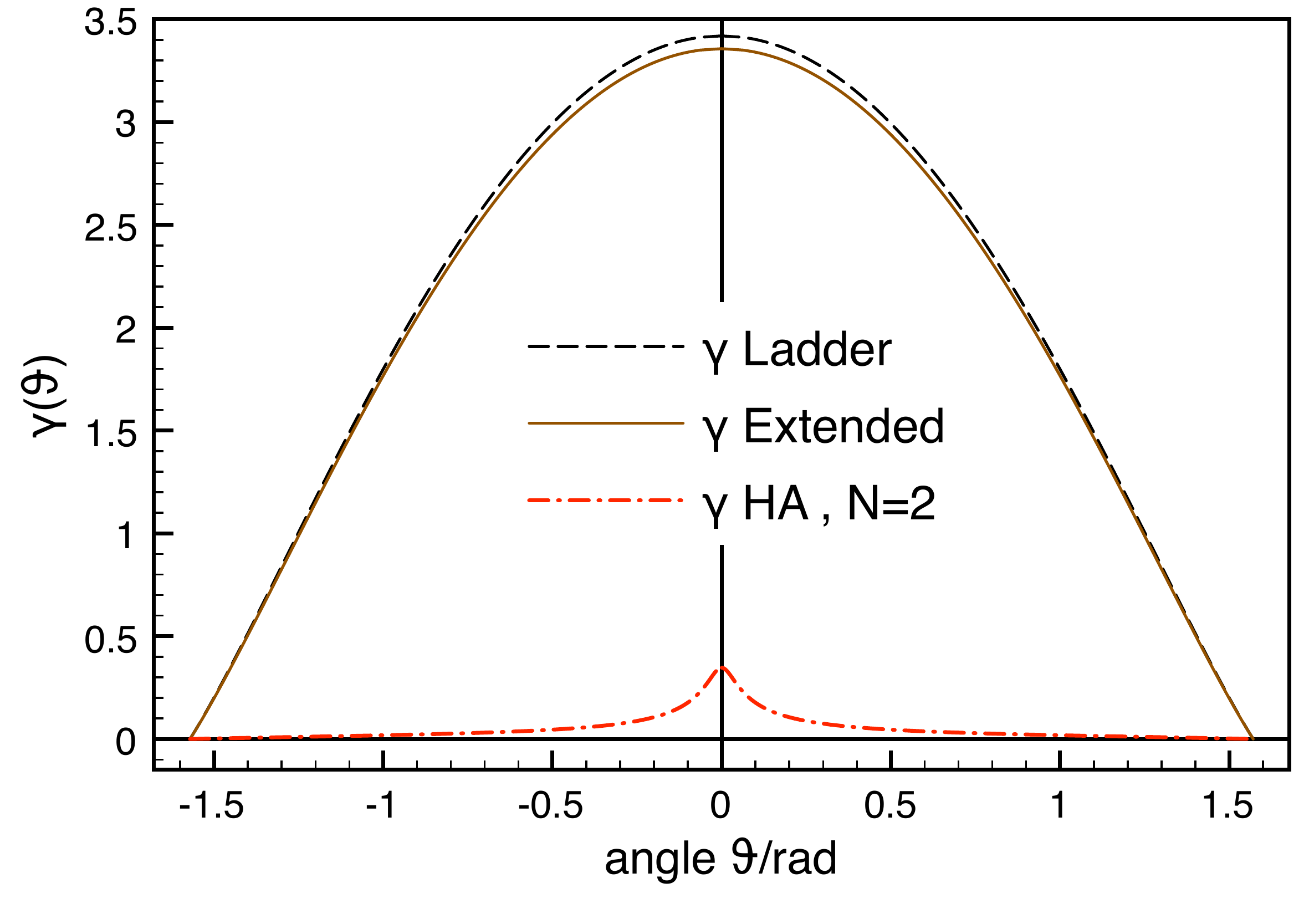}}
\subfigure[]{\includegraphics[height=63mm]{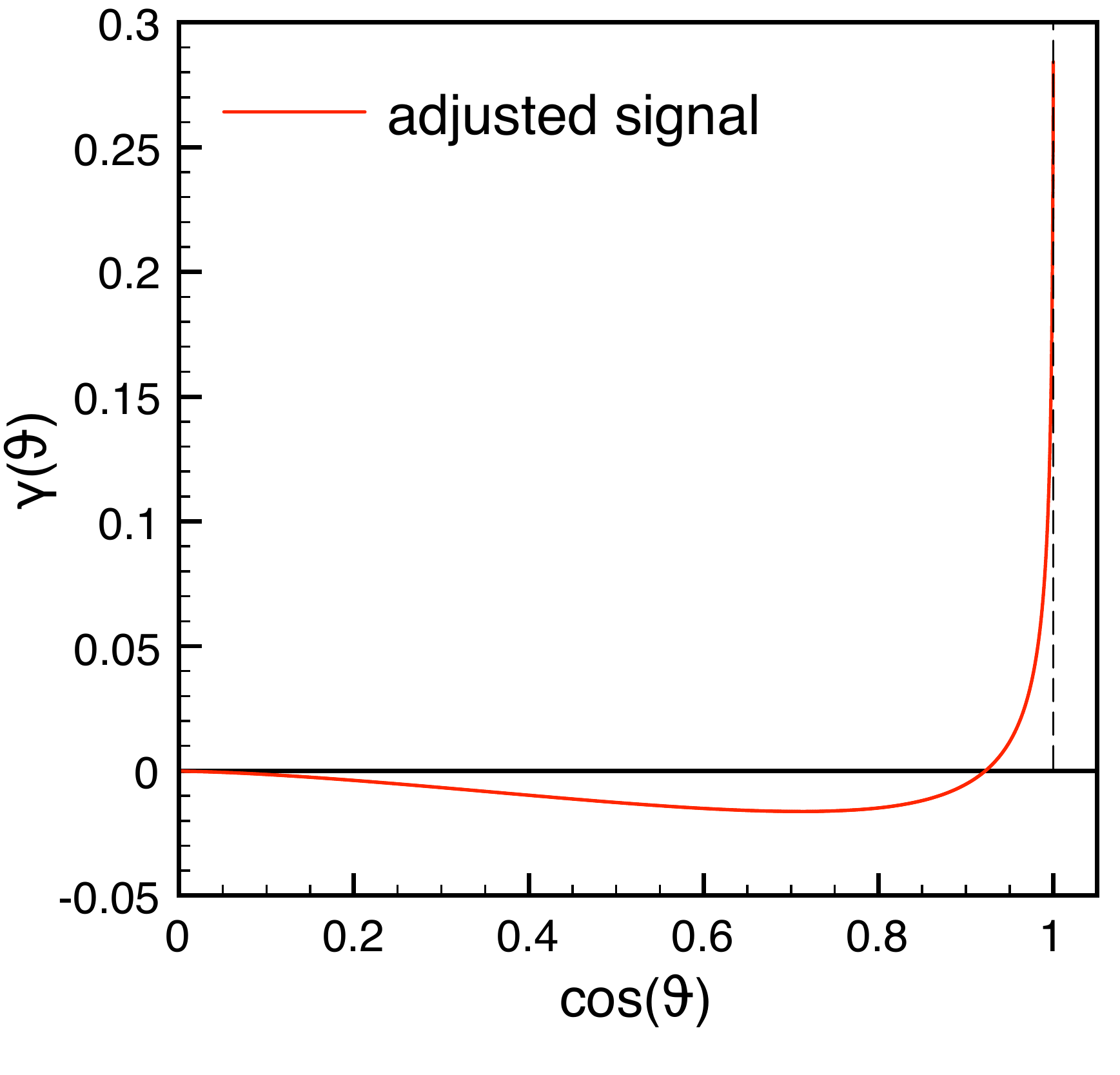}}
\caption{(a) Comparison of the incoherent background distribution obtained when only ladder propagation is considered [$\gamma_{Ladder}(\vartheta)$] with the case where the ladder approximation has been extended 
[$\gamma_{Extended}(\vartheta)$, see eq.~(\ref{eq:gammaextended})] by the $H_{A,N=2}$-loop processes shown in figure~\ref{ext}.
 System parameters: $k\ell=30$ and $b=10$. For the $H_{A, N=2}$-scattering process we also show the corresponding backscattering peak [$\gamma_{H_{A, N=2}}(\vartheta)$]. Whereas adding this $\gamma_{H_{A, N=2}}$-peak on top of pure ladder propagation $\gamma_{Ladder}$ would violate flux conservation, the latter is restored when replacing  $\gamma_{Ladder}$ by the slightly reduced $\gamma_{Extended}$. (b) Comparison of the new signal (backscattering peak on top of adjusted background) to the old, unaltered background. Here we plot $(\gamma_{H_{A, N=2}}+\gamma_{Extended}-\gamma_{Ladder})$ as a function of $\cos\theta$,
 in order to verify that the total integral over $\cos\theta$ vanishes.
}
\label{ALL}
\end{figure}

Finally, figure \ref{ALL} shows the solution of eq.~(\ref{eq:gammaN2tot}) for $k\ell=30$ and $b=10$, obtained by Monte-Carlo simulation of an anisotropic random walk as explained above. As expected, the renormalized background $\gamma_{Extended}(\vartheta)$ is slightly reduced with respect to the ladder background such that, as we have checked, the sum of $\gamma_{Extended}(\vartheta)$ and the
double scattering cone $\gamma_{H_{A, N=2}}(\vartheta)$ fulfills flux conservation:
$\int_{-1}^{+1}\frac{d\cos\vartheta}{2}\left(\gamma_{H_{A, N=2}}(\vartheta)+\gamma_{Extended}(\vartheta)\right)=1$.

Let us conclude this section with some remarks concerning the coherent backscattering enhancement factor \cite{wiersma95}.
Note that figure~\ref{ALL} only displays the double scattering contribution to coherent backscattering, whereas the renormalized background $\gamma_{Extended}$ contains all scattering orders. Adding further scattering orders to the coherent backscattering cone,
e.g. taking into also $\gamma_{H_{A,N>2}}$ or any other of the processes shown in figure~\ref{scatt}, requires a further renormalization of the background, scaling like $\ln(k\ell)/(k\ell)^2$ for $k\ell\to\infty$, in order to ensure flux conservation. We expect that this renormalization can be performed in the same way as demonstrated here for $\gamma_{H_{A,N=2}}(\theta)$ (see also the corresponding discussion in the conclusion below). Note, however, that for each of the processes contributing to the background $\gamma_{Extended}$ in figure~\ref{ALL}, a corresponding crossed diagram -- giving rise to an identical contribution in exact backscattering direction $\theta=0$ -- can be found by reversing one of the amplitudes. Therefore, the total coherent backscattering enhancement factor remains unchanged up to the order $1/(k\ell)$, and deviations from the ideal value two (after subtracting single scattering from the background) scale at least like $\ln(k\ell)/(k\ell)^2$ for our model of a white noise Gaussian random potential (and we expect that the same holds true also for Gaussian potentials with non-vanishing correlation length). Note that this does not contradict the result of \cite{wiersma95}, where deviations of the backscattering enhancement factor of the order $1/(k\ell)$ have been shown to occur as a consequence of recurrent scattering for a discrete scatterer model (which is non-Gaussian, since recurrent scattering amounts to a non-vanishing fourth order cumulant).

\section{Conclusion and outlook}
Concerning the mechanism restoring energy conservation in the effect of coherent backscattering, we have obtained  the following results:
The leading order of coherent backscattering, found to arise from $H_{A,N=2}$-scattering (double scattering event of the crossed contribution), cannot be counterbalanced by scattering processes including one additional scatterer, since the contributions of these scattering processes where found to scale in higher order of $1/(k\ell)$. We could not establish any cancellations between the higher order processes, either, which contradicts the previous theory presented in~\cite{Fieb}.

On the other hand, it turned out to be crucial to develop a full description of coherent backscattering in the frame of the loop propagation processes as considered in the description of weak localisation in infinite disordered media~\cite{Pub}. Considering all possible constellations of a weak localisation loop with respect to the boundary surface of the scattering medium, one finds additional scattering contributions which equally have to be taken into account. For the leading order contribution (scaling like $1/(k\ell)$ for $k\ell\to\infty$) we present an approximate description of the scattering process as an anisotropic random walk. Within the frame of this approximation, it is possible to obtain a flux-conserving description of coherent backscattering. For the remaining higher order processes depicted in figure~\ref{scatt}, we raise the hypothesis that each contribution to coherent backscattering is in the same way cancelled intrinsically in the frame of a complete description accounting for all possible constellations in the presence of a boundary surface: According to this hypothesis, flux conservation is established for an arbitrary contribution
 to the
irreducible intensity vertex  $U(\mathbf{r}_1,\mathbf{r}_2,\mathbf{r}_3,\mathbf{r}_4)$ (e.g. $U_{H_{A,N=2}}(\mathbf{r}_1,\mathbf{r}_2,\mathbf{r}_3,\mathbf{r}_4)=\delta(\mathbf{r}_1-\mathbf{r}_4)\delta(\mathbf{r}_2-\mathbf{r}_3)u^4 \left|\overline{G}(|\mathbf{r}_1-\mathbf{r}_2|)\right|^2$ in figure~\ref{ext}) when taking into account all four possibilities of connecting $(\mathbf{r}_1,\mathbf{r}_2)$ and $(\mathbf{r}_3,\mathbf{r}_4)$ either directly to the incoming (or outgoing) wave outside the scattering medium or to other points ($\mathbf{r}_s$ or $\mathbf{r}_f$ in figure~\ref{ext}) within the scattering medium.

A possible ansatz in order to prove this hypothesis might be a generalization of what is known as the \textit{Ward identity}~\cite{vollhardt80}, ensuring flux conservation for scattering processes in infinite disordered media, to the case of a non-translationally invariant scattering medium in presence of a boundary surface.

From an aesthetical point of view, it is certainly satisfying to dispose of a theory of coherent backscattering where flux conservation is intrinsically built in. This is the main motivation of the present paper. Practical consequences concerning, e.g., the comparison between  theoretical and experimental coherent backscattering cones in order to determine properties of the scattering medium, remain to be investigated. In this case, effects like internal reflections due to a refractive index mismatch \cite{zhu91} or the finite correlation length of the random medium  -- which have been neglected in the present paper -- 
must also be taken into account.
It will be interesting to see to what extent the flux-conserving scattering diagrams identified in the present paper then lead to predictions that differ from other approaches (e.g. \cite{Fieb}), and whether this will enable a more accurate determination of scattering properties such as the transport mean free path.

\ack
We would like to thank Felix Eckert for helpful discussion and many a friendly word as well as Reinhard Knothe for answers and support with several bigger or smaller programming troubles.

\end{document}